\documentclass[preprint2]{aastex6}

\usepackage{graphicx}
\usepackage[FIGTOPCAP]{subfigure}
\usepackage{amsmath,booktabs,threeparttable,url, bm}
\usepackage[hyphenbreaks]{breakurl}
\usepackage{CJKutf8}

\newcommand{\cntext}[1]{\begin{CJK*}{UTF8}{bsmi}#1\end{CJK*}}

\usepackage{natbib}
\begin{document}
\title{Equation of State Dependent Dynamics and Multimessenger Signals from Stellar-mass Black Hole Formation}

\author{Kuo-Chuan Pan (\cntext{潘國全})$^{1,2}$, Matthias Liebend\"{o}rfer$^{3}$,
Sean~M.~Couch$^{1,2,4,5}$ and Friedrich-Karl Thielemann$^{3}$}
\affil{$^1$Department of Physics and Astronomy, Michigan State University, East Lansing, MI 48824, USA; pankuoch@msu.edu}
\affil{$^2$Joint Institute for Nuclear Astrophysics, Michigan State University, East Lansing, MI 48824, USA}
\affil{$^3$Departement Physik, Universit\"{a}t Basel, Klingelbergstrasse 82, CH-4056 Basel, Switzerland}
\affil{$^4$Department of Computational Mathematics, Science, and Engineering, Michigan State University, East Lansing, MI 48824, USA}
\affil{$^5$National Superconducting Cyclotron Laboratory, Michigan State University, East Lansing, MI 48824, USA}


\begin{abstract}

We investigate axisymmetric black hole~(BH) formation and its gravitational wave (GW) and neutrino signals 
with self-consistent core-collapse supernova simulations of a non-rotating $40~M_\odot$ progenitor star
using the isotropic diffusion source approximation for the neutrino transport 
and a modified gravitational potential for general relativistic effects.
We consider four different neutron star (NS) equations of state~(EoS):
LS220, SFHo, BHB$\Lambda\phi$ and DD2,
and study the impact of the EoS on BH formation dynamics and GW emission.
We find that the BH formation time is sensitive to the EoS from 460 to $>1300$~ms and
is delayed in multiple dimensions for $\sim~100-250$~ms due to the finite entropy effects.
Depending on the EoS, our simulations show the possibility that shock revival 
can occur along with the collapse of the proto-neutron star~(PNS) to a BH.
The gravitational waveforms contain four major features 
that are similar to previous studies but show extreme values:
(1)~a low frequency signal ($\sim~300-500$~Hz) from core-bounce and prompt convection,
(2)~a strong signal from the PNS g-mode oscillation among other features,
(3)~a high frequency signal from the PNS inner-core convection, and
(4)~signals from the standing accretion shock instability and convection.
The peak frequency at the onset of BH formation reaches to $\sim~2.3$~kHz.
The characteristic amplitude of a 10~kpc object at peak frequency is detectable but close to the
noise threshold of the Advanced~LIGO and KAGRA,
suggesting that the next generation gravitational wave detector will need to improve the sensitivity at the kHz domain
to better observe stellar-mass BH formation from core-collapse supernovae or failed supernovae.
\end{abstract}



\section{INTRODUCTION}

Supernovae are the spectacular birth places of neutron stars (NSs)
and stellar-mass black holes (BHs) in the universe.
Stellar-mass BHs can be formed either by accretion onto a NS,
or by a failed SN, where the stalled bounce shock never revives
\citep{2003ApJ...591..288H, 2008ApJ...684.1336K}.
Optical observations of a disappearing star seem to confirm the existence of failed SNe
\citep{2015MNRAS.450.3289G, 2017MNRAS.468.4968A}.

Although there is no firm conclusion on the explodability of core-collapse SN from first principle calculations,
one-dimensional (1D) spherical symmetry simulations of failed SN and BH formation have been investigated by
\cite{2004ApJS..150..263L, 2011ApJ...730...70O, 2007ApJ...667..382S, 2012ApJ...757...69U, 
2015ApJ...809..116C, 2016ApJ...821...38S}.
These studies suggest that the mass range of SN progenitors that may end up with a BH is not monotonic in mass,
and the BH formation time is sensitive to the progenitor density structure
(or the so-called compactness parameter introduced in \citealt{2011ApJ...730...70O})
and the nuclear equation of state (EoS) used.
However, spherically symmetric simulations cannot accurately account for multi-dimensional effects such
as turbulence, convection, and rotation, which are considered crucial ingredients for achieving successful explosions.

Spherically symmetric simulations with a simplified description of rotation and/or artificial heating (piston, thermal bomb, or PUSH) 
\citep{1995ApJS..101..181W, 1996ApJ...460..408T, 2015ApJ...806..275P, 2016ApJ...818..124E}, 
successfully reproduce several observables, 
such as nickel mass production ($\sim 0.1 M_\odot$) and explosion energies ($\sim 10^{51}$~erg).
However, these simulations have to be calibrated from multi-dimensional simulations.
In addition, the prediction of gravitational wave (GW) emission requires the calculation of the mass quadrupole moment terms
from multi-dimensional simulations.

\cite{2005PhRvD..71h4013S} and \cite{2011PhRvL.106p1103O}  have performed
two-dimensional (2D) and three-dimensional (3D) general relativistic (GR) simulations of rotating stars
from core collapse to BH formation with a polytropic EoS and a parameterized adiabatic index $\Gamma$.
As the transport of neutrinos was ignored,
these simulations end up with prompt BH formation within $\sim 150$~ms,
which is a much shorter time than in 1D simulations ($> 500$~ms) with neutrino transport
and microphysical EoSs from \cite{ls220} in \cite{2011ApJ...730...70O}.
Neutrino transport and a nuclear EoS are essential ingredients in supernova simulations
that are driven by the neutrino-driven mechanism
(see recent reviews in \citealt{2012ARNPS..62..407J, 2013RvMP...85..245B, 2016NCimR..39....1M, 2016ARNPS..66..341J}
and references therein).

Recently, a series of 2D and 3D simulations with neutrino transport and microphysical EoS have been performed,
and GW emission has been investigated by
\cite{2008A&A...490..231S, 2012PhRvD..86b4026O, 2013ApJ...766...43M, 2015PhRvD..92h4040Y, 2016ApJ...829L..14K, 2017MNRAS.468.2032A};
;\cite{2017arXiv170107325Y}; and \cite{2017ApJ...851...62K}.
However, due to the extensive computational time of these simulations with neutrino transport,
most of these studies only simulated the first few hundred milliseconds post-bounce
during which BHs do not have enough time to form.
\cite{2013ApJ...779L..18C} performed a 2D {\tt CoCoNut} GR simulation of a rapidly rotating star
with a neutrino leakage scheme and with the Lattimer \& Swesty EoS (with the incompressibility $K=220$~MeV, LS220, \citealt{ls220}).
They found a very strong GW signal from the violent proto-neutron star (PNS) dynamics and observed a BH formed
at $\sim 1.6$~s post-bounce.

While rotating models show a very strong and robust bounce signal
in the GW emission that is dominated by the rotational speed \citep{2017PhRvD..95f3019R},
in this paper, we consider non-rotating models only and focus on the impact of the NS EoS on the shock dynamics,
neutrino emission, and GW signatures. 
The NS EoS, which describes the mass-radius relationship of a NS, 
is still unclear \citep{2017RvMP...89a5007O}.
Our prediction of multi-messenger signals with different EoS can be used 
as numerical constraints for future observations.

This paper is organized as follows.
In Section~\ref{sec_method}, we describe our simulation code, numerical methods, 
physics involved, and initial conditions.
In Section~\ref{sec_bh}, we present our simulation results and describe the general evolution 
from core collapse and bounce to BH formation.
The evolution of the PNS, and its growth to BH is described in Section~\ref{sec_pns}.
In Section~\ref{sec_gw}, we show the GW signals from our axisymmetric simulations.
We discuss and investigate the impact of the EoS on the shock dynamics, neutrino emissions,
and GW signals in Section~\ref{sec_discussion}.
Finally, we summarize our results and conclude in Section~\ref{sec_conclusions}.

\section{NUMERICAL METHODS} \label{sec_method}

\begin{figure*}
\epsscale{1.2}
\plotone{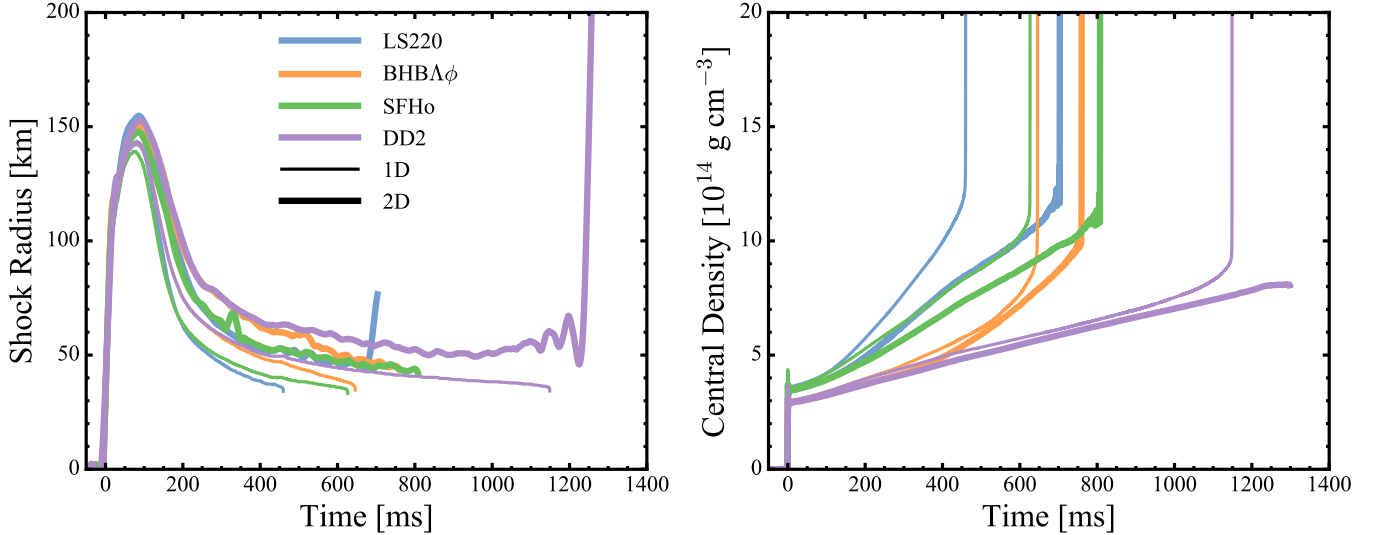}
\caption{\label{fig_rsh}
Time evolution of averaged shock radius (left) and central density (right).
Different colors represent simulations with different nuclear EoS.
Thick and thin lines indicate simulations in 2D and 1D, respectively.}
\end{figure*}

We use {\tt FLASH}\footnote{\url{http://flash.uchicago.edu}} version~4 \citep{2000ApJS..131..273F, 2008PhST..132a4046D}
to solve the Newtonian multi-dimensional neutrino radiation-hydrodynamics
with the Isotropic Diffusion Source Approximation (IDSA; \citealt{idsa})
for the transport of electron flavor neutrinos.
Heavy neutrinos are described by a leakage scheme
\citep{1998ApJ...507..339H, 2003MNRAS.342..673R}.

In IDSA, the distribution function of transported neutrinos is decomposed into a free streaming and a trapped component.
It is assumed that the evolution of these two components can be solved separately and linked by a diffusion source term \citep{idsa}.
Spherically symmetric simulations with the IDSA have shown good agreement with the Boltzmann transport simulations \citep{idsa}.
For multi-dimensional simulations, one could either implement the IDSA in a ``ray-by-ray'' approach 
\citep{2012ApJ...749...98T, 2013ApJ...764...99S},
or solve the diffusion source term in multiple dimensions 
but keep the streaming component spherically symmetric \citep{idsa2d, 2017nuco.confb0703P}.
In this paper, we use the latter approach for our 2D simulations.

The new multipole Poisson solver \citep{2013ApJ...778..181C}
with a maximum multipole value $l_{\rm max} = 16$ is used for the calculation of self-gravity.
We modify the monopole moment of the gravitational potential
to effectively include the GR effects \citep{2018ApJ...854...63O}
based on the Case A implementation that is described in \cite{2006A&A...445..273M}.

The general grid setup is similar to what has been implemented and described by
\cite{2013ApJ...765...29C}, \cite{2018ApJ...854...63O} and \cite{idsa2d}.
The simulation box includes the inner $10^4$~km of the progenitor in 1D spherical coordinates or 2D cylindrical coordinates. 
In order to approximate the stellar envelope, a power-law profile in spherical radius is
used as the boundary condition for density and velocity at the outer edges of our computational domain.
Although the free fall time scale from our boundary ($t_{\rm ff} \sim 1.7~s$) is longer 
than our simulation time, the standard ``outflow'' boundary condition will overestimate 
the mass accretion flow at late times \citep{2013ApJ...765...29C}.
We employ nine levels of adaptive mesh refinement (AMR) in our simulations.
The central $r \lesssim 120$~km sphere has the smallest zone width of $0.488$~km
and the AMR level decreases based on the distance to the center,
giving an effective angular resolution of $0.2^\circ - 0.4^\circ$.
We use 20 neutrino energy bins that are spaced logarithmically from 3 to 300 MeV for the electron flavor neutrinos
and the IDSA solver has been enhanced with GPU acceleration with OpenACC \citep{2017nuco.confb0703P}.
A single node simulation with a Nvidia P100 GPU on the Cray XC50 {\tt Piz-Daint} supercomputer 
at the Swiss National Supercomputing Centre (CSCS) is 2.4 times faster than the same run without GPU.
We performed self-consistent simulations from stellar core collapse, bounce, post-bounce evolution, to BH formation.

We use the 40 solar mass progenitor with solar metallicity (s40) from \cite{2007PhR...442..269W}.
Based on the 1D BH formation study in \cite{2011ApJ...730...70O},
the s40 progenitor has the shortest BH formation time among the progenitors in \cite{2007PhR...442..269W},
which is a convenient choice to save computing time in multi-dimensional simulations.
 
As inelastic scattering processes, which are important during the collapse, have been ignored in our IDSA solver, 
we use IDSA to update the neutrino quantities while a parameterized deleptonization (PD) scheme \citep{2005ApJ...633.1042L} is used
to update the electron fraction ($Y_e$), entropy and the momentum transfer from neutrino pressure during collapse.
The PD parameters\footnote{$\rho_1= 4 \times 10^8$~g~cm$^{-3}$, $\rho_2= 6 \times 10^{12}$~g~cm$^{-3}$,
$Y_1= 0.5$, $Y_2=0.015$, and $Y_c=0.272$} are calibrated based on the bounce profile of
an {\tt Agile-Boltztran} \citep{2004ApJS..150..263L} simulation with the LS220 EoS.
In principle, we should calibrate the PD parameters for each EoS, considering that the bounce profile is EoS dependent.
We use the same PD parameters for all different EoS for consistency and simplicity. 
Different PD parameters give slightly different bounce time and PNS core structure, 
but the post-bounce evolution is not very sensitive to PD parameters at late times. 
Core bounce is defined when the central density is above $2\times 10^{14}$~g~cm$^{-3}$ 
and the maximum core entropy reaches 3~$k_{\rm B}$~baryon$^{-1}$.
After bounce, we turn off the PD module and use the IDSA solver for the rest of the calculation.

The nuclear EoS unit in {\tt FLASH}, which incorporates the finite-temperature EoS routines
from \cite{2010CQGra..27k4103O, 2013ApJ...765...29C}, is used.
We consider four different nuclear EoS to study the impact of the EoS on the dynamics of BH formation
and GW signals, including the LS220 EoS,
the Banik, Hempel, \& Bandyopadhyay EoS (BHB$\Lambda\phi$) \citep{bhblp},
the Steiner, Fischer, \& Hempel (SFHo) EoS \citep{sfho},
and the Hempel \& Schaffner-Bielich (HS) DD2 EoS \citep{dd2}.

The LS220 EoS is based on the single nucleus approximation for heavy nuclei
and does not fill the constraints from chiral effective field theory \citep{2013PhRvC..88b5802K, 2014EPJA...50...46F}.
However, the LS220 EoS is one of the most common EoS in the supernova community and has been widely used in many supernova simulations.
Therefore, we include the LS220 EoS as a reference.
On the other hand, the DD2 EoS uses the density-dependent relativistic mean-field (RMF) interactions of
\cite{2010PhRvC..81a5803T}.
\cite{2015PhRvC..91d5805H} have found a good agreement of cluster formation with nuclear experiments in the DD2 EoS,
but the NS radius with DD2 EoS is inconsistent with the observations by \citep{2013ApJ...765L...5S}.
The SFHo EoS has similar nuclear properties to the DD2 EoS, but the mass-radius is tuned
to fit the NS radius observation \citep{2013ApJ...765L...5S}.
On top of the DD2 EoS, the BHB$\Lambda\phi$ EoS additionally includes $\Lambda$ hyperons
and hyperon-hyperon interactions mediated by $\phi$ mesons.
The $\Lambda$ hyperon makes the EoS softer
and therefore could accelerate the formation of a black hole \citep{2015ApJ...809..116C}.
Without the hyperon-hyperon interaction, the BHB$\Lambda$ EoS fails to produce NS with mass $M > 2M_\odot$ \citep{bhblp}.
Hence, the $2 M_\odot$ NS observation by \cite{2013Sci...340..448A} gives a strong constraint on the nuclear EoS.

We approximate nuclear burning by assuming nuclear statistical equilibrium (NSE), 
which is accurate for all regions that experience temperatures beyond $\sim 5\times 10^9$~K.

\begin{figure*}
\epsscale{1.2}
\plotone{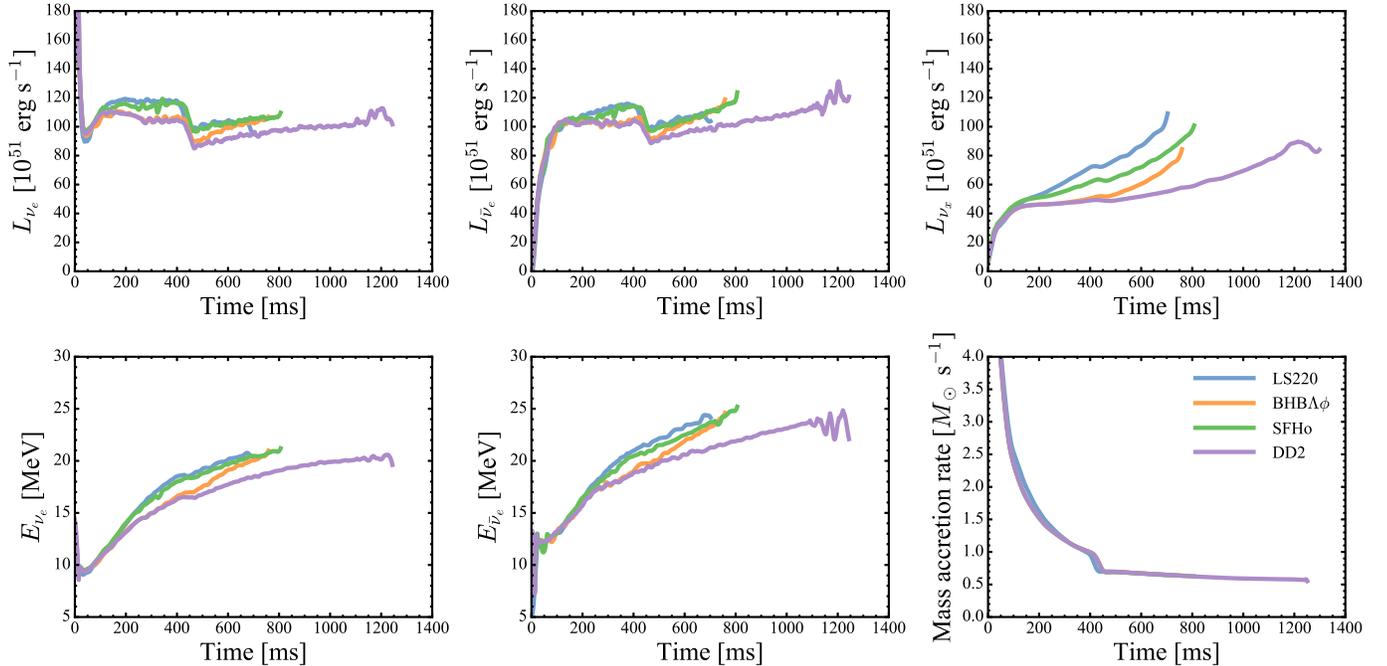}
\caption{\label{fig_lnu}
The left (middle) panels show neutrino luminosity and mean energy of electron flavor neutrino (anti-neutrino),
the upper right panel represents the luminosity of one representative heavy neutrino species,
and the lower right panel indicates the mass accretion rate (measured at $r=500$~km) as functions of time for our 2D runs.
Different colors represent simulations with different nuclear EoS.
}
\end{figure*}

\begin{deluxetable}{lcrrrrr}
\tabletypesize{\scriptsize}
\tablecaption{Black Hole Formation Properties \label{tab_bh}}
\tablehead{
\colhead{EoS} & \colhead{Dim.} & \colhead{$t_{\rm BH}$} & \colhead{$M_{\rm b, BH}$} & \colhead{$f_{\rm peak, BH}$} & \colhead{$N_{\nu_e}$} & \colhead{$N_{\bar{\nu}_e}$}\\
\colhead{} &\colhead{} & \colhead{(ms)} & \colhead{($M_\odot$)} & \colhead{(kHz)} & \colhead{($10^{57}$)} & \colhead{($10^{57}$)}
}
\startdata
LS220 & 1D & 460 & 2.22 & --- & 2.57 & 1.74\\
BHB$\Lambda\phi$ & 1D & 646 & 2.33 & --- & 2.91 & 2.26 \\
SFHo & 1D & 626 & 2.32 & --- & 2.86 & 2.23\\
DD2 & 1D & 1149 & 2.56 & --- & 4.35 & 3.61\\
LS220 & 2D & 704 & 2.53 & 2.3 & 3.45 & 2.38 \\
BHB$\Lambda\phi$ & 2D & 760 & 2.56 & 2.2 & 3.44 & 2.58\\
SFHo & 2D & 808 & 2.59 & 2.3 & 3.60 & 2.73\\
DD2 & 2D & $> 1300$ & $>2.86$ & $>2.2$ & $>5.11$ & $>4.15$
\enddata
\tablecomments{$t_{\rm BH}$ is the BH formation time in millisecond post-bounce;
$M_{\rm b, BH}$ is the maximum baryonic PNS mass right before BH formation;
$f_{\rm peak, BH}$ is the peak GW frequency from the g-mode PNS oscillation;
and $N_{\nu_e}$ and $N_{\bar{\nu}_e}$ are the total number emission of electron neutrinos and anti-neutrinos, respectively.}
\end{deluxetable}

\section{From Core Bounce to BH Formation} \label{sec_bh}

Becasue we consider non-rotating models, 2D runs during the collapse phase show identical evolution as in 1D.
Core bounce occurs at $368$~ms in all simulations except for the runs with LS220 EoS (at $466$~ms).
The $\sim100$~ms delay in the LS220 runs leads to a slightly different mass accretion history in the post-bounce phase.
After bounce, a shock is launched and expands to $\sim140-160$~km at $\sim 80$~ms post-bounce.
2D runs have a $\sim 10\%$ larger shock radius than 1D runs due to the amplification 
by prompt convection seeded by grid perturbations \citep{2017nuco.confb0703P}.

Figure~\ref{fig_rsh} shows the evolution of the average shock radii and central densities as functions of time.
The LS220 and SFHo EoS runs have a slightly higher central density at bounce than the BHB$\Lambda\phi$ and DD2 EoS.
The BHB$\Lambda\phi$ and DD2 runs show identical evolutions in the first 300~ms,
since the hyperon effects are not important when the density is lower than $5\times 10^{14}$~g~cm$^{-3}$.

The prompt convection sets in at around 20~ms post-bounce,
letting the 2D runs deviate from the 1Ds by having a slightly larger maximum shock radius at about 100~ms post-bounce.
The standing accretion shock instability (SASI; \citealt{2003ApJ...584..971B}) starts to develop after $\sim150$~ms post-bounce,
enlarging the shock radius in 2D.
Among our simulations, the 1D LS220 run reaches the second collapse first at $\sim 460$~ms post-bounce
and the calculations are stopped when the central density reaches the density limit in the EoS table ($\rho_{\rm max} \sim 7 \times 10^{15}$~g~cm$^{-3}$).
In other words, a BH is formed.
The second collapse can be seen as a sudden increase of the central density in Figure~\ref{fig_rsh}.

Table~\ref{tab_bh} summarizes the BH formation properties in all of our 1D and 2D simulations.
All of our simulations result in BH formation except the 2D DD2 run.
The 2D DD2 run explodes at $\sim 1.27$~s post-bounce.
The explosion time is defined when the averaged shock radius exceeds 400~km and never returns during the simulation.
Once a star explodes, there are still accretion funnels that continue adding to the PNS mass,
but the central density stops increasing after tens of milliseconds (the thick purple line in Figure~\ref{fig_rsh}).
At this point, the total baryon mass of the PNS, $M_{\rm PNS} =2.86 M_\odot$,
exceeds the mass limit of the DD2 EoS for a cold NS ($M_{\rm max}= 2.42 M_\odot$, \citealt{dd2}).
Therefore, once the PNS cools down, a BH will form.
We terminate the 2D DD2 run when its averaged shock radius reaches 1000~km.
In the 2D LS220 run, a BH is formed at $704$~ms post-bounce. At the same time, the shock starts to revive,
suggesting that a successful explosion together with a BH formation is possible without the need of fallback accretion.

\begin{figure}
\epsscale{1.2}
\plotone{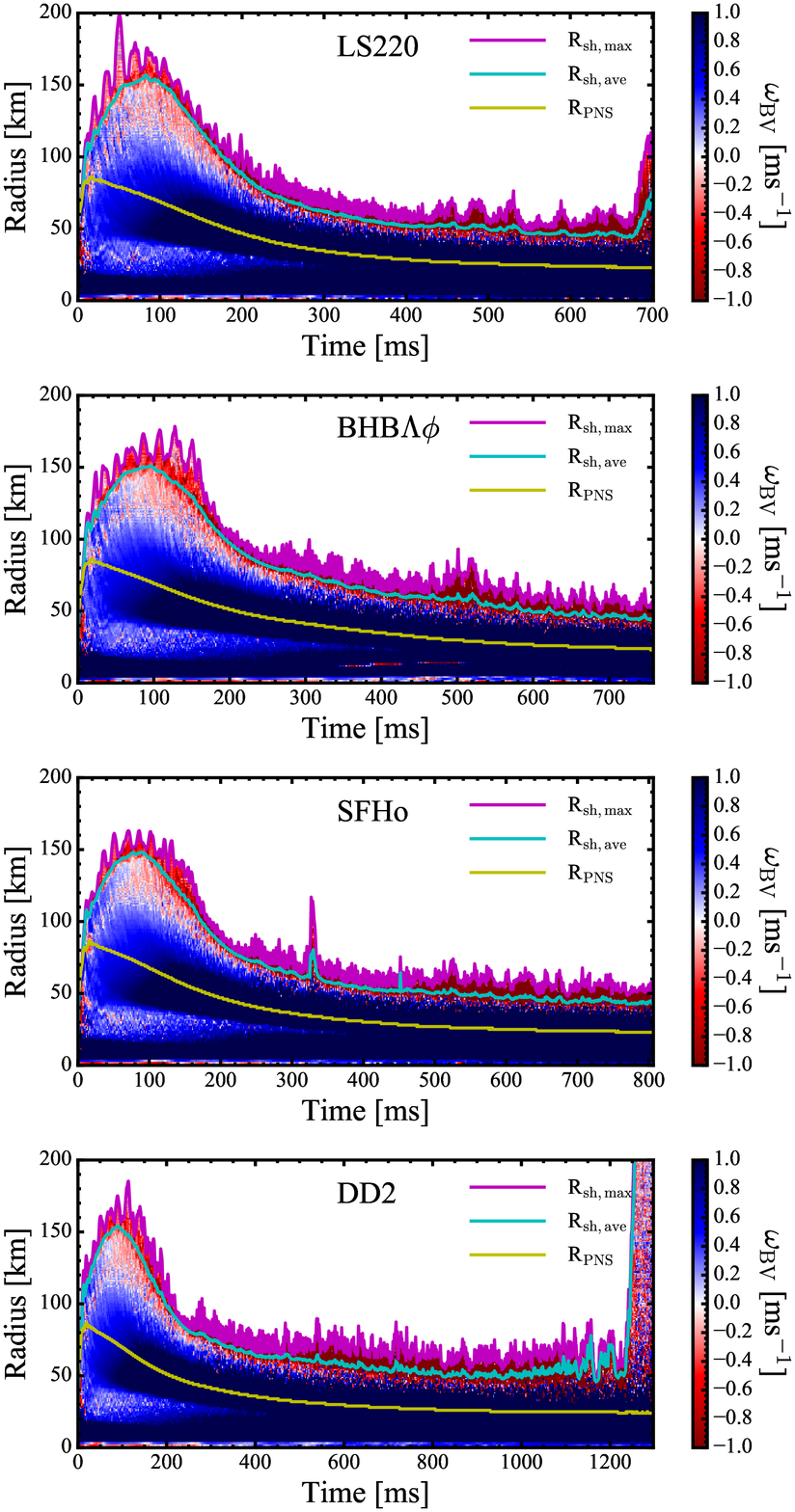}
\caption{\label{fig_wbv}
Color maps of the Brunt-V\"{a}is\"{a}l\"{a} frequency for different EoS as functions of time and radius.
The red color represents the unstable regions based on the Ledoux criterion.
The magenta and light blue lines show the maximum and averaged shock radius.
The yellow line indicates the radius of the PNS.
}
\end{figure}

\begin{figure}
\epsscale{1.2}
\plotone{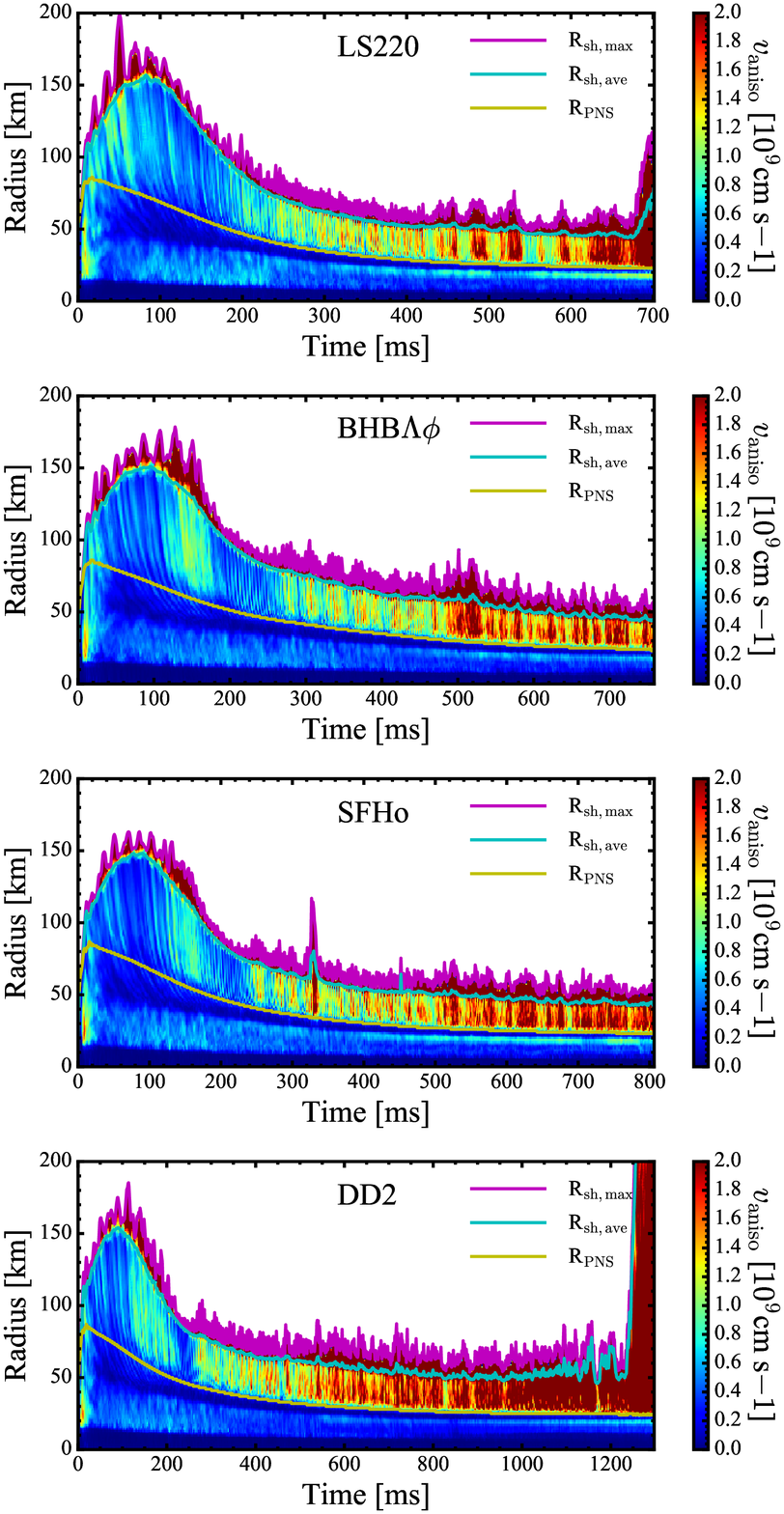}
\caption{\label{fig_vaniso}
Same as Figure~\ref{fig_wbv} but for the anisotropic velocities (defined in Equation~\ref{eq_vaniso}).
}
\end{figure}

Figure~\ref{fig_lnu} shows the neutrino luminosity and mean energy as functions of time.
The neutrino signals in our simulations can be categorized into two groups:
(1)  LS220/SFHo runs, and (2) BHB$\Lambda\phi$/DD2 runs.
In the first hundred milliseconds post-bounce, the neutrino luminosity and mean energy are similar for all EoS.
After about $\sim 150$~ms post-bounce, two groups start to develop.
The neutrino luminosity and mean energy in the LS220 and SFHo EoS runs grow faster than in
the results with BHB$\Lambda\phi$ and DD2 EoS.
When the Si/O shell interface reaches the accretion shock at $\sim 420$~ms post-bounce,
there is a $\sim 20~\%$ drop of the neutrino luminosity, likely due to the decrease of the mass accretion rate.
After $420$~ms, the electron neutrino and electron anti-neutrino luminosity increase linearly in time.
The LS220 and SFHo EoS show slightly higher neutrino luminosities and mean energies.
This is likely due to the more compact and denser PNS core in these two EoS (see Figure~\ref{fig_rsh}).
Note that our $\mu / \tau$ neutrinos are described by a leakage scheme, which 
does not take into account redshift effects.
Therefore, the high $\mu / \tau$ luminosity at the time close to BH formation might be overestimated.
The M1 transport simulation in \cite{2015ApJS..219...24O} shows a much flatter $\mu / \tau$ luminosity at late times.
Although the neutrino luminosity and mean energy show only a small difference between different EoS,
the significant difference of the BH formation time leads to a very different total neutrino number emission (see Table~\ref{tab_bh}),
which can be a useful diagnostic for future neutrino observations.

The main difference between 2D and 1D runs is convection in 2D.
There are two regions showing strong convection:  
(1) the PNS, and
(2) the gain region.
These two convective regions can be understood from a local stability analysis,
using the Ledoux criterion \citep{1947ApJ...105..305L},
\begin{equation}
C_L = -\left( \frac{\partial \rho}{\partial p}\right)_{s,Y_l}\left[ \left( \frac{\partial p}{\partial s} \right)_{\rho,Y_l} \left( \frac{d s}{d r} \right) +  \left( \frac{\partial p}{\partial Y_l} \right)_{\rho, s} \left( \frac{d Y_l}{d r} \right) \right],
\end{equation}
where $\rho$ is the density, $p$ is the pressure, $s$ is the entropy, $Y_l$ is the lepton fraction,
and $r$ is the distance from the center of the PNS.
We approximate $Y_l \sim Y_e$ for convenience.
Following the description in \cite{2006A&A...447.1049B},
the Brunt-V\"{a}is\"{a}l\"{a} frequency, $\omega_{BV}$, which describes the linear growth frequency for convection,
can be written by,
\begin{equation}
\omega_{BV} = {\rm sign}(C_L) \sqrt{\left| \frac{C_L}{\rho} \frac{d \Phi}{d r} \right|},
\end{equation}
where $\Phi$ is the local gravitational potential and an approximation of $d\Phi/dr \sim -GM(r)/r^{2}$ is used.

Figure~\ref{fig_wbv} shows the evolution of the angle-averaged Brunt-V\"{a}is\"{a}l\"{a} frequency from our four 2D runs.
The yellow lines trace the evolution of the PNS radius (defined by $\rho = 10^{11}$~g~cm$^{-3}$).
One can clearly see that prompt convection is occurring at around $50$~km at $\sim 20$~ms post-bounce.
After $\sim 400$~ms when the shock radius shrinks,
the region of PNS convection becomes smaller and is limited to a small region close to the PNS surface.

In addition to the Brunt-V\"{a}is\"{a}l\"{a} frequency, 
another useful quantity to investigate fluid instability and convection is the so-called anisotropic velocity $v_{\rm aniso}$
that is defined in \cite{2012ApJ...749...98T} as,
\begin{equation}
\label{eq_vaniso}
v_{\rm aniso} = \sqrt{\frac{\left< \rho \left[ (v_r - \left< v_r \right>_{4\pi})^2 + v_\phi^2 \right] \right>_{4 \pi}}{\left< \rho \right>_{4\pi}}},
\end{equation}
where $\left< v_r\right>_{4\pi}$ and $\left<\rho\right>_{4\pi}$ are the spherically averaged radial velocity and density, respectively.
$v_\phi$ is the tangential velocity.
The angle-averaged anisotropic velocity evolution from our four 2D simulations is shown in Figure~\ref{fig_vaniso}.
Strong anisotropic velocities in the two convective regions dramatically perturb the PNS and will contribute to the GW emission.

%

\begin{figure*}
\epsscale{0.5}
\plotone{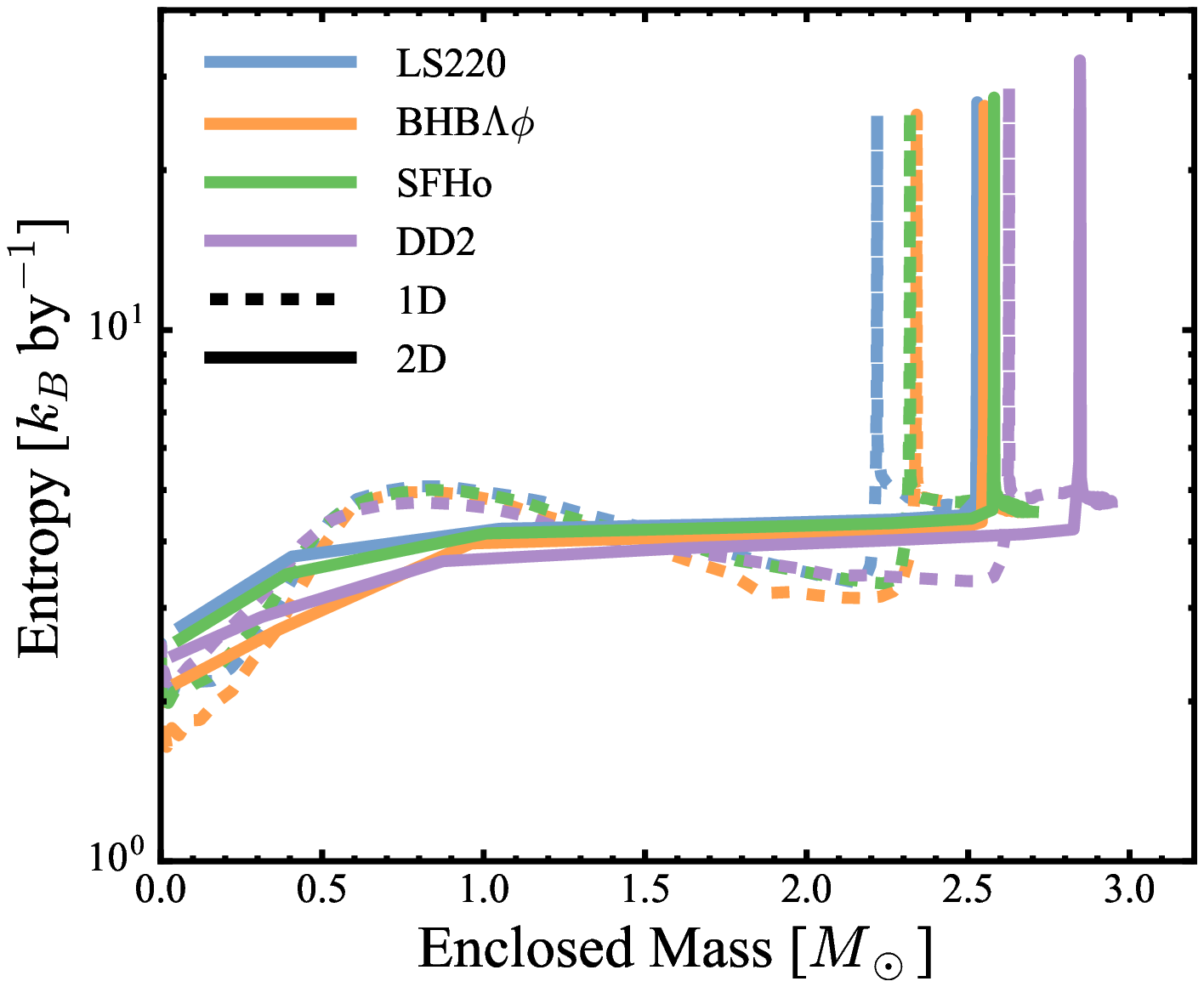}
\plotone{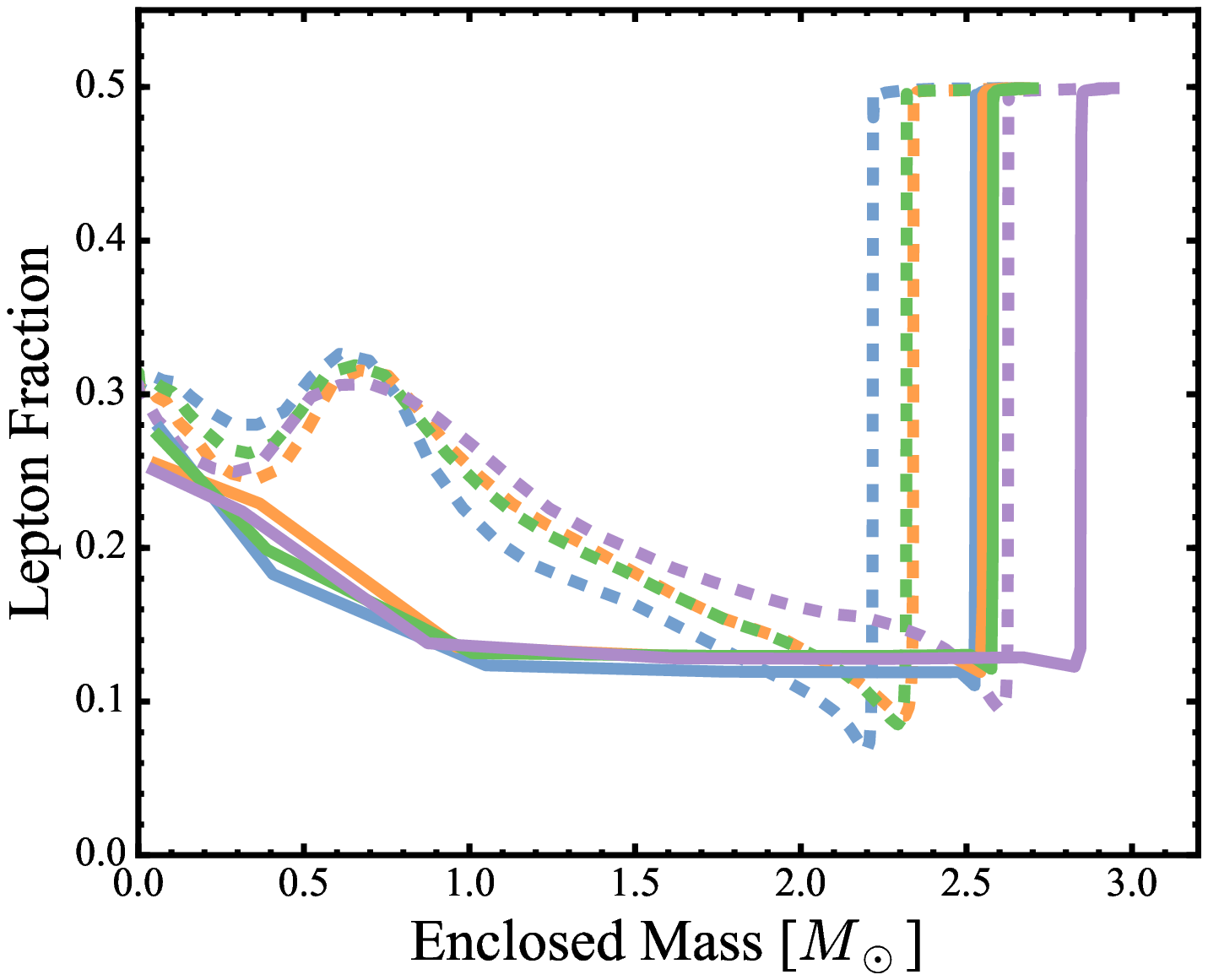}
\caption{\label{fig_entr}
Angle-averaged entropy and lepton fraction as functions of enclosed mass.
Solid lines show the profiles from 2D simulations and dashed lines are 1D.
All profiles are plotted at 10~ms before BH formation, except the 2D DD2 run (at $t_{\rm pb} = 1.245$~s)
which explodes before BH formation.
}
\end{figure*}

\section{PNS Evolution} \label{sec_pns}

Convection does not only redistribute matter within the PNS but also enlarges the maximum stable PNS mass.
Figure~\ref{fig_entr} shows the angle-averaged entropy and lepton fraction profiles
at $10$~ms before BH formation (except the 2D DD2 run).
The high-entropy peaks correspond to the gain region where neutrino heating is strong.
Because the mass in the gain region is small ($M_{\rm gain} \lesssim 0.001M_\odot$),
the PNS mass can be approximated by the enclosed mass at these high-entropy peaks in Figure~\ref{fig_entr}.
For 1D runs, there is a negative entropy gradient at around $M \sim 0.5 M_\odot$ which drives the PNS convection.
2D runs show relatively flat entropy profiles in the region of $1 M_\odot <  M < 2.5 M_\odot$.
All 2D runs have a higher PNS mass than in 1D.

Here, we approximate the lepton fraction by $Y_l = Y_e + Y_{\nu_e}^t - Y_{\bar{\nu}_e}^t$,
where $Y_{\nu_e}^t$ and $Y_{\bar{\nu}_e}^t$ are trapped electron neutrino and trapped electron anti-neutrino fractions.
Similar to the entropy profiles, the negative composition gradient in 1D has been flattened in 2D runs,
suggesting that this increases the PNS mass in 2D due to convection.
The increment of the PNS mass in 2D is mainly due to convection.
In addition, convection also affects the neutrino heating/cooling, giving a higher PNS radius in 2D.
The magnitude of this effect depends on the EoS (see Table~\ref{tab_bh}).

%
\begin{figure*}
\epsscale{1.1}
\plotone{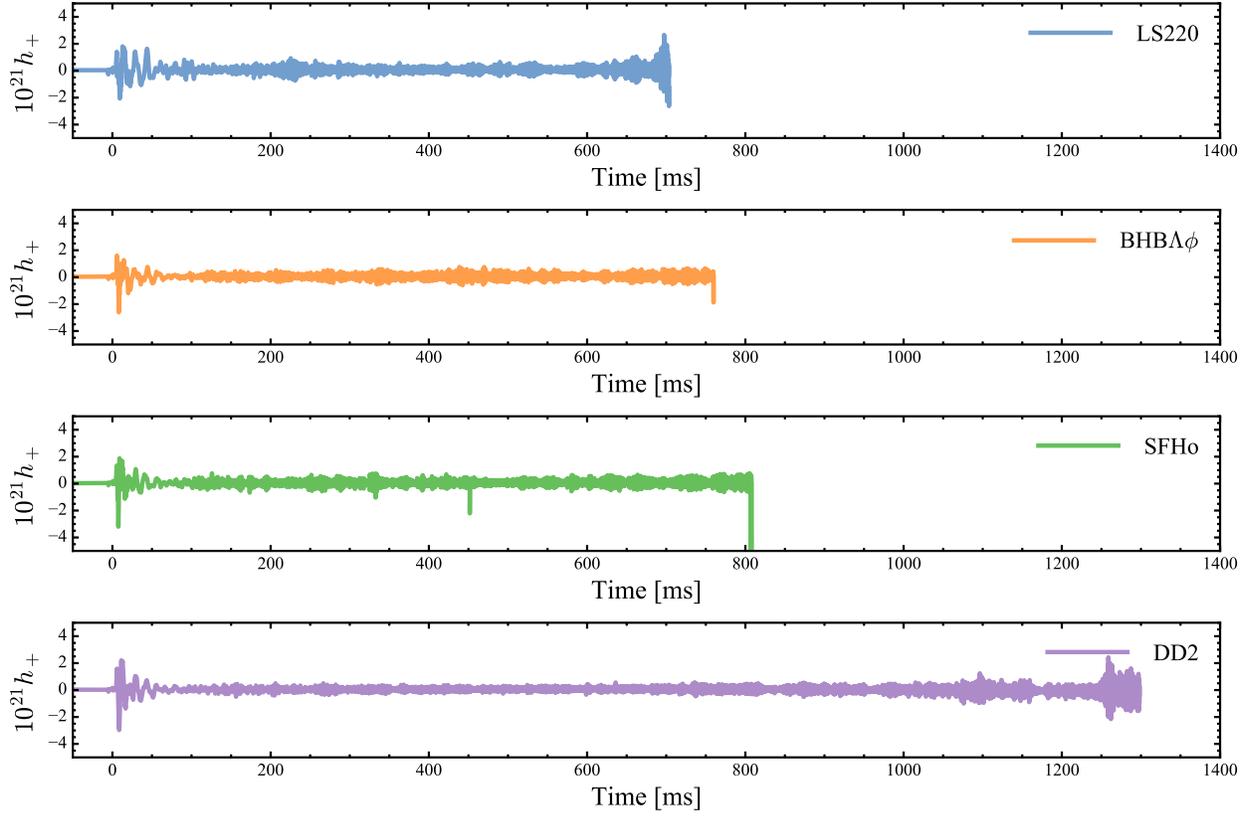}
\caption{\label{fig_gw}
GW amplitudes as functions of time after core bounce with an assumption of $10$~kpc distance to the source.
From top to bottom, each panel shows 2D simulations with LS220, BHB$\Lambda\phi$, SFHo, and DD2 EoS, respectively.
In the top three panels, a BH is formed at the end of the simulation.
}
\end{figure*}
\begin{figure*}
\epsscale{1.22}
\plotone{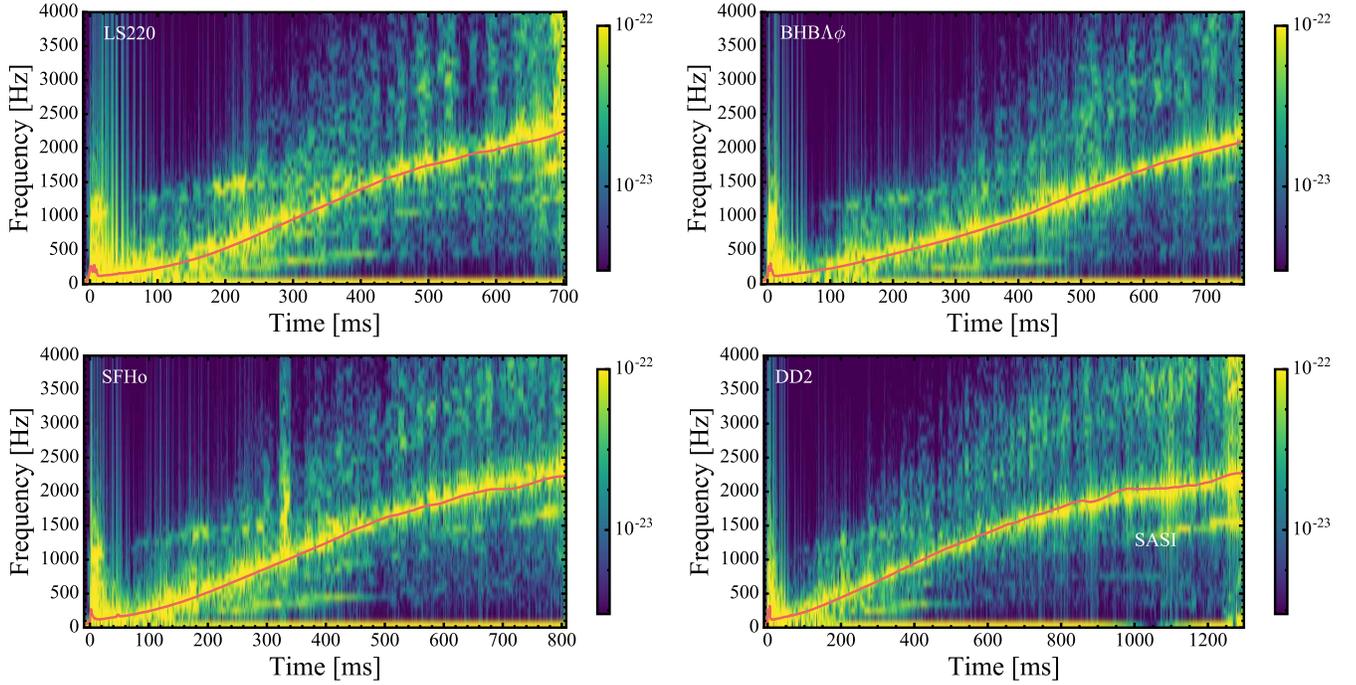}
\caption{\label{fig_spectrogram}
GW amplitude spectrograms for a window of 10~ms with different EoS, assuming a distance of 10~kpc and looking from the equator ($\sin \theta=1$).
The colorbars are plotted in a logarithmic scale.
The red lines are plotted from the analytical formula describing the PNS g-mode oscillation as in Equation~\ref{eq_g_mode}.
}
\end{figure*}

\section{GW Signals} \label{sec_gw}

We extract the GW signals based on the mass quadrupole formula using the approximation
\begin{equation}
h_+ \approx \frac{3}{2}\frac{G}{D c^4}\ddot{{\boldsymbol I}}_{zz}  \sin^2\theta,
\end{equation}
where $D$ is the distance to the source, ${\boldsymbol I}$ is the mass quadrupole tensor,
and $\theta$ is the inclination angle
\citep{1997PThPS.128..183O, 2008A&A...490..231S, 2009ApJ...707.1173M}.
Figure~\ref{fig_gw} shows the gravitational waveforms of our four 2D runs, assuming a distance of 10~kpc and $\theta=\pi/2$.
The general features are similar to what has been described in \cite{2009ApJ...707.1173M} in the context of non-exploding models:
a strong signal occurs during the first 50~ms due to the core bounce and prompt convection.
After $\sim 50$~ms, the frequency rises but the amplitude decreases in time until the SASI sets in.
For the 2D LS220 and 2D DD2 runs, high-amplitude GW are emitted when the shock is revived.
A strong, short ($< 1$ms), and high-frequency signal is also observed when the PNS collapses to a BH.

To better understand the GW frequency and time evolution,
we show in Figure~\ref{fig_spectrogram} the GW amplitude spectrogram by performing
a short-time Fourier transform with a time window of $10$~ms.
The peak frequency in yellow stems from the well-known PNS g-mode oscillation
and can be characterized by a combination of PNS mass $M_{\rm PNS}$ and radius $R_{\rm PNS}$
\citep{2009ApJ...707.1173M, 2013ApJ...766...43M, 2013ApJ...779L..18C, 2016PhRvD..94d4043S}.
We believe that the peak frequency from the g-mode oscillation at the surface of the PNS can be fit by
\begin{equation}
\label{eq_g_mode}
f_{\rm peak} \sim \frac{1}{2\pi} \frac{GM_{\rm PNS}}{R_{\rm PNS}^2c} \sqrt{1.1 \frac{m_n}{\langle E_{\bar{\nu}_e} \rangle}}. 
\end{equation}
Note that Equation~\ref{eq_g_mode} is adapted from Equation~17 in \cite{2013ApJ...766...43M}
but without the factor of $ \left(1-\frac{GM}{Rc^2}\right)^2$.
We find that Equation~\ref{eq_g_mode} fits our data better (see the red lines in Figure~\ref{fig_spectrogram}).
The difference might come from our treatment of the effective GR potential 
and the lack of redshift corrections in the IDSA.
We will discuss this difference more in the next section.
The peak frequency from g-mode PNS oscillations at BH formation reaches about $2.2 \sim 2.3$~kHz 
(see Table~\ref{tab_bh}).

From core bounce to about three hundred millisecond post-bounce,
a high-frequency component, starting from $\sim 1000$~Hz to $\sim 1500$~Hz,
is associated with the PNS convection of the inner core at $r \sim 30$~km,
as the perturbation frequency of the quadrupole moment is closely related to the sound speed
and can be seen in Figure~\ref{fig_wbv} and \ref{fig_vaniso} as well.
In addition, there are several components associated with the SASI in the convective gain region
that let high-density clumps fall onto the PNS 
(see discussion in \citealt{2013ApJ...779L..18C}).

It should be noted that Equation~\ref{eq_g_mode} contains three major physical quantities: 
$M_{\rm PNS}$, $R_{\rm PNS}$, and $\langle E_{\bar{\nu}_e} \rangle$, 
where $M_{\rm PNS}$, $R_{\rm PNS}$ describe the mass-radius relationship of a given EoS. 
Therefore, measuring both GW and neutrino emissions 
would give observational constraints on NS EoS. 

%
\section{DISCUSSION} \label{sec_discussion}
\begin{figure}
\epsscale{1.22}
\plotone{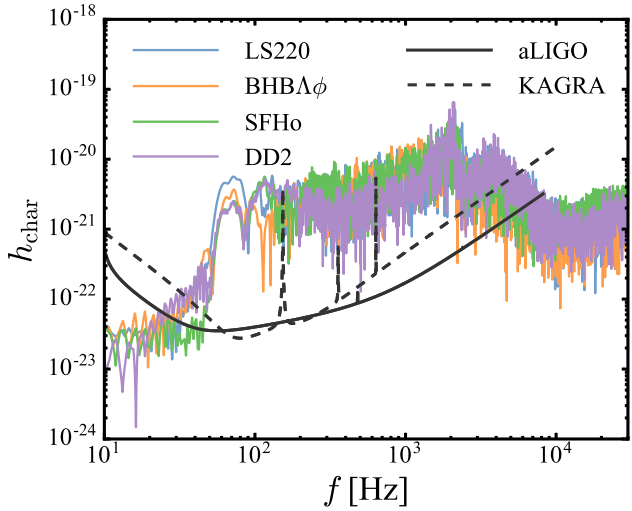}
\caption{\label{fig_spectra}
The characteristic GW amplitudes, $h_{\rm char}$, for the four EoS.
The solid (dashed) gray line is the approximate noise thresholds for Advanced LIGO (KAGRA) at $10$~kpc.
}
\end{figure}
\cite{2005PhRvD..71h4013S} and \cite{2011PhRvL.106p1103O} performed BH formation simulations with full GR in multiple dimensions
and investigated the GW emission.
In both studies, a polytropic EoS was used and neutrino transport was ignored,
leading to unrealistically short BH formation times ($t_{\rm BH} < 200$~ms).
In \cite{2011PhRvL.106p1103O}, the non-rotating model has a very weak GW signal and is excluded from their analysis.
On the other hand, the rotating models show strong GW signals and the peak frequency at BH formation is about $\sim 2.9 - 3.9$~kHz,
depending on the rotational speed,
where the most slowly rotating model has the shortest BH formation time and the highest peak frequency.
As the GW peak frequency from the g-mode oscillation is strongly associated with the PNS mass and radius (i.e., the NS EoS),
their peak frequency might be overestimated.

Recently, \cite{2013ApJ...779L..18C} investigated BH formation from a rotating collapsar with the LS220 EoS and full GR.
Their setup is more realistic and closer to ours.
Unlike our investigation, they use another progenitor star ($35 M_\odot$) and include rotation.
The less massive progenitor and rotation causes the BH formation to be delayed to $\sim 1.6$~s,
but the overall shock dynamics and GW signals are consistent with our results.
\cite{2013ApJ...779L..18C} use the {\tt CoCoNuT} code to solve the relativistic hydrodynamics with the XCFC approximation \citep{2009PhRvD..79b4017C},
which is physically more accurate than the effective GR potential \citep{2006A&A...445..273M} used in our simulations.
Nevertheless, their peak frequency from the PNS g-mode oscillation is about 2~kHz, similar to ours.
As discussed in Section~\ref{sec_gw}, a factor of $(1-GM/rc^2)^2$ is not included in our fitted peak frequency,
implying that our peak frequencies may be slightly overestimated due to the effective GR potential and lack of redshift corrections in the IDSA.

Our GW signatures from the SASI motions are also consistent with the results in \cite{2013ApJ...779L..18C}.
It should be noted that both the simulation by \cite{2013ApJ...779L..18C} and our simulations are 2D simulations.
The SASI motion will be very different in 3D due to the additional degree of freedom.
Therefore, these GW signals from the SASI need to be validated by future 3D simulations.

To detect the GW emissions from core-collapse supernovae or failed supernovae,
it is useful to evaluate the characteristic amplitude $h_{\rm char}$.
We follow \cite{2009ApJ...707.1173M} to calculate the dimensionless characteristic amplitude \citep{1998PhRvD..57.4535F},
\begin{equation}
h_{\rm char} = \frac{1}{D} \sqrt{\frac{2G}{\pi^2c^3}\frac{dE_{GW}}{df}},
\end{equation}
where $dE_{GW}/df$ is the GW spectral energy density,
\begin{equation}
\frac{dE_{GW}}{df}= \frac{3G}{5c^5}\left( 2\pi f \right)^2 \arrowvert \tilde{A} \arrowvert^2,
\end{equation}
and $\tilde{A}$ is the Fourier transform of $A \equiv \ddot{I}_{zz}$ computed by
\begin{equation}
\tilde{A} (f) = \int_{-\infty}^\infty A(t) e^{- 2\pi i f t} dt.
\end{equation}
Figure~\ref{fig_spectra} shows the characteristic amplitude as function of frequency for our four 2D runs,
assuming a distance of 10~kpc.
The sensitivity curves of the Advanced LIGO (aLIGO) and KAGRA are also plotted for a comparison
\citep{2010CQGra..27h4006H, 2010CQGra..27h4004K}.
It is crucial that the low-frequency window ($\sim 60$~Hz $< 1000$~Hz) is detectable with the aLIGO and KAGRA
and that differences due to the EoS are distinguishable (see Figure~\ref{fig_spectra}).
However, although the GW emissions from the high-frequency window ($f > 1000$~Hz) have the largest characteristic amplitude,
the signal strain is very close to the limit of the aLIGO and KAGRA, making it difficult to detect with the current GW detectors.
Therefore, in order to confirm BH formations in core-collapse supernovae or failed supernovae,
the next-generation GW detectors will need to improve the sensitivity in the $\sim 1000$~Hz window.

%
\section{SUMMARY AND CONCLUSIONS} \label{sec_conclusions}

We have performed axisymmetric 2D core-collapse supernova simulations of the non-rotating s40 progenitor \citep{2007PhR...442..269W}
with IDSA neutrino transport and a modified GR potential.
We find that the BH formation time is very sensitive to the NS EoS and varies from $\sim450$~ms to $>1300$~ms.
For a given EoS, the 2D simulation has a larger PNS mass before BH formation than the 1D counterpart
due to finite temperature effects and convection,
leading both to have a longer BH formation time.
With the BHB$\Lambda\phi$ and SFHo EoS in 2D, a BH is formed at 760~ms and 810~ms, respectively,
and no shock revival before BH formation has been found.
Therefore, no optical or very weak optical emissions from BH accretion are expected in these failed supernovae \citep{2017MNRAS.468.4968A}.
On the other hand, the 2D LS220 and DD2 runs end up with BH formation and shock revival in the same simulation 
without the need of fallback accretion at late time.

Due to the delay of BH formation in 2D, the total number of neutrinos emitted will be higher in multiple dimensions
than in the corresponding 1D simulations, e.g.~\cite{2011ApJ...730...70O} (see Table~\ref{tab_bh}).
The strongest GW emissions stem from the g-mode PNS oscillations, 
but additional features from the SASI and PNS inner-core convective are still visible 
in the $\sim 100 - 2000$~Hz window, which is possible to be detected by aLIGO and KAGRA.

Our predictions of neutrino and GW emissions suggest that
BH formations in nearby ($d \lesssim 10$~kpc) core-collapse supernovae or failed supernovae can be detected by
the current neutrino and GW detectors such as aLIGO, KAGRA, and Super-Kamiokande,
but the GW peak frequency at BH formation ($f_{\rm peak} \sim 2$~kHz) is close to the sensitivity threshold.

\acknowledgments
We acknowledge valuable conversations and input from Matthias Hempel, Takami Kuroda, and Evan O'Connor
on the supernova equations of state and gravitational wave emissions.
This work was supported by the European Research Council (ERC; FP7)
under ERC Advanced Grant Agreement N$^\circ$~321263-FISH,
by the Swiss National Science Foundation (SNF),
by the Swiss Platform for Advanced Scientific Computing (PASC) project DIAPHANE.
The Basel group is a member of the COST Action New Compstar.
S.M.C. is supported in part by the U.S. Department of Energy, Office of Science under award DE-SC0015904.
{\tt FLASH} was in part developed by the DOE NNSA-ASC OASCR Flash Center at the University of Chicago.
The simulations have been carried out at the CSCS ({\tt Piz-Daint}) under grant No.~661
and on the high performance computing center (HPCC) at Michigan State University.
Analysis and visualization of simulation data were completed using the analysis toolkit {\tt yt}
\citep{2011ApJS..192....9T}.

%

\end{document}